\documentclass[12pt]{article}
\usepackage{color,multicol}
\usepackage{amsmath}
\usepackage{amssymb}
\usepackage{epsfig}

\newcommand{\set}[1]{{\mathbb{#1}}}

\begin{document}

\title{Elasticity theory of structuring\footnote{Original version 23 April 2013. Journal-ref: Risk, December (2016), pp.\! 81-86.\vspace*{1mm}}}
\author{Andrei N. Soklakov\footnote{Head of Strategic Development, Asia-Pacific Equities, Deutsche Bank.\vspace*{1mm}\newline
{\sl The views expressed herein should not be considered as investment advice or promotion. They represent personal research of the author and do not necessarily reflect the view of his employers, or their associates or affiliates.} Andrei.Soklakov@(db.com, gmail.com).}}
\date{}
\maketitle

\begin{center}
\parbox{14cm}{
{\small
Financial derivatives have often been criticized as casino-style betting instruments. It turns out that many naive ways of making them are indeed equivalent to gambling. Fortunately, this inadvertent effect can be understood and prevented. We present a theory of product design which achieves that.
} }
\end{center}

\section{Introduction}

In~\cite{Soklakov_2014_WQS} we reviewed the shortcomings of the pre-crisis approach to product design and advocated the need for a better, more quantitative, approach. We started to build this approach in~\cite{Soklakov_2011} and provided further illustrations in~\cite{Soklakov_2013a}.

So far, most of our examples have been centered around the important special case of the growth-optimizing investor. This investor defines growth in terms of the compounded rate of return (logarithmic rate of return) and seeks to maximize the expectation of this rate regardless of the risks. First introduced by Bernoulli in 1738, this is probably the oldest and one of the best researched benchmarks of investment behavior~\cite{Bernoulli_1738, Christensen_2005}.

With the growth-optimizing (very risky) investor on the one hand and the risk-free (bond) investor on the other hand, we begun to understand the range of products which most sensible investors would want to use~\cite{Soklakov_2011}. In this paper we refine our framework allowing investors to control their risk appetite in a more precise way.

Combining Bayesian information processing with rational optimization gives us a simple new tool for product design -- the payoff elasticity equation. This is the main technical result of the paper. After some demonstrations of the equation, including detection and prevention of inadvertent gambling, we summarize our approach to product design as a self-contained manufacturing process. In terms of infrastructure, this is fully compatible with existing business environments.\\

\section{Deriving derivatives}

Let $x$ be a random variable with some financial significance -- this can be anything from a stock price or an average commodity value to temperature readings in Texas. The variable $x$ can be a vector, containing several such examples as components. Vector-valued $x$ can also be used to hold time series of a market variable (see~\cite{Soklakov_2013a} for a concrete example). Financial product is called a derivative of $x$ if its payoff $F$ is defined as a function of $x$.

Behavioral Finance teaches us to be very careful when using intuition in finance. In terms of payoff functions, this means that not every ad-hoc proposal for $F$ is good; not even when it is clear and well-defined. Superstitions and compulsive disorders often manifest themselves as transparent, well-defined procedures which lead people to disastrous results. The best constructive way of supporting human intuition is to use the scientific approach. In the case of financial derivatives it begins with making sure that investment structuring is guided by the logical laws of information processing.

\section{Investor equivalence principle}

In this presentation we consider the simplest case of a real-valued underlying, $x\in \set{R}$. This assumption is not crucial in any way and we use it purely for notational clarity. Following~\cite{Soklakov_2011}, we partition the range of possible values of $x$ into non-overlapping intervals using a discrete mesh $(\dots,x_i,x_{i+1},\dots)$. Imagine now a set of securities $\{s_i\}$, where each $s_i$ pays 1 when~$x$ fixes between $x_i$ and $x_{i+1}$ and zero otherwise. We can get a quote for purchasing the securities, $p_i={\rm quoted\ price}(s_i)$. It is tempting to call $\{s_i\}$ the Arrow-Debreu securities and connect (undiscounted) $\{p_i\}$ to the risk-neutral probabilities for~$x$. Let us resist this temptation as it would force us to make a lot of unnecessary assumptions. At this point we do not require a two-way liquid arbitrage-free market of securities for each distinct state of the economy. All we need is someone who would be willing to sell $\{s_i\}$ -- a set which may contain just a couple of securities.

In~\cite{Soklakov_2011} we introduced $\{s_i\}$ as binary spreads and argued that for a very large class of investors the problem of optimal investment is equivalent to optimal splitting of capital across the $\{s_i\}$.

Let $\{\beta_i\}$ be the proportions in which the investor decides to partition their capital ($\sum_i\beta_i=1$). Only one security among $\{s_i\}$ can mature in the money, so the payoff $F$ from the investment is easy to compute
\begin{equation} \label{Eq:Fk}
    F_k=\beta_k r_k\,,
\end{equation}
where $k$ is the index of the security that matures in the money and $r_k=1/p_k$ is the quoted return on that security. We define the market-implied distribution by normalizing the prices $m_k=p_k/\sum_i p_i$ and rewrite the above equation in the \lq\lq Bayesian" form
\begin{equation} \label{Eq:Bayes_General}
    \beta_k=F_k\, m_k\,,
\end{equation}
where we decided not to burden our notation with trivial normalization constants (this can be arranged formally by re-defining $F_k$, or by leaving $m_k$ not normalized).

Equation~(\ref{Eq:Bayes_General}) covers all investment scenarios which can be broken down into mutually exclusive events $\{s_i\}$. Given that the choice of the underlying variable remains completely up to the investor this is a very general setting. Indeed, not only can we use $x$ to mean anything we want (from exotic strategies to weather readings), the choice of $\beta_k$ in (\ref{Eq:Fk}) is also very general -- so far we didn't even assume the investor to be rational in any way. Equation~(\ref{Eq:Bayes_General}) is just basic accounting. In the following we use Eq.~(\ref{Eq:Bayes_General}) as a property which defines {\emph{any investor}}.

Let $\{b_k\}$ be investor-believed probabilities that the corresponding securities $\{s_k\}$ mature in the money. In~\cite{Soklakov_2011} we considered a special case of a growth-optimizing investor. In this special case the fractions $\{\beta_k\}$ coincide with $\{b_k\}$~\cite{Kelly_1956}. Equation~(\ref{Eq:Bayes_General}) becomes
\begin{equation}\label{Eq:Bayes_GrowthOptimal}
b_k=f_k\, m_k\,,
\end{equation}
where $f_k$ denotes the growth-optimal payoff. We revisit this case in more detail later.

Although growth-optimizing investors form a very specific group, we see that focusing on this group does not at all reduce the range of possible investment decisions. Indeed, both $\{\beta_i\}$ and $\{b_i\}$ must add up to one, but with no further constraints, Eqs.~(\ref{Eq:Bayes_General}) and (\ref{Eq:Bayes_GrowthOptimal}) are solved by the same set of payoffs, $\{f\}=\{F\}$. By pairing up these payoffs we can now make an even more detailed observation, which, in view of its generality, we formulate here as a principle:
\begin{center}
{\bf \emph{Any investor} can be viewed as growth-optimizing.}
\end{center}
More precisely, general investor chooses the same product, $F=f$, i.e. behaves in the same way, as some growth-optimizing investor whose beliefs~$\{b_k\}$ happen to coincide with~$\{\beta_k\}$.

The principle focuses our attention on the actions of the investor -- their net economic behavior. The logical integrity of these actions can be checked in many ways. The principle suggests doing this from the point of view of a growth-optimizing investor -- a very convenient choice.

\section{Logical investors}
In~\cite{Soklakov_2011} we made a connection between Eq.~(\ref{Eq:Bayes_GrowthOptimal}) and Bayes' theorem. We recognized the market-implied $\{m_k\}$ and the investor-believed $\{b_k\}$ as the prior and the posterior distributions and highlighted the fundamental connection between the optimal payoff structures and the likelihood functions. This gave us some understanding of how the growth-optimizing investors learn and then express their knowledge through trading.

In the previous section we also discovered that, in terms of their net economic behavior, the growth-optimizing investors are surprisingly versatile. It looks like they can be educated to act as if they had sophisticated risk preferences. The aim of this section is to describe and understand this education process mathematically. To this end we narrow the scope of our investigations and focus on the case of rational investors.

It can be shown that rational investors behave as if they were maximizing the expected value of utility~\cite{NeumannMorgenstern_1944}. The expected utility approach which follows from this observation is well known in economic theory. We will use it here as well. It is important, however, to emphasize that by adopting the expected utility approach we do not reduce the generality of our arguments in the same way as it happens in economic theories. For us, rationality is not an assumption -- it is part of the goal. We understand that being rational is not easy, so we build tools which facilitate rational behavior.

Consider a rational investor with utility $u()$ and a view on $x$ given by the probabilities $\{b_i\}$, where each $b_i$ measures the degree of investor's belief for $x$ to end up between $x_i$ and $x_{i+1}$. Because the logarithm is a monotonically increasing function, we can, without loss of generality, write the utility as a function of the logarithmic rate of return. We also allow the utility to depend on $x$ explicitly. The optimal investment is found by maximizing the expectation
\begin{equation}\label{Eq:ExpectedUtility}
    \sum_ib_iu\Big(\ln(\beta_i r_i),x_i\Big)
\end{equation}
over all possible proportions $\{\beta_i\}$ subject to the budget constraint $\sum_i\beta_i=1$. The Lagrangian for this optimization reads
\begin{equation}\label{Eq:Lagrangian}
    {\cal L}(\{\beta_i\},\lambda)=\sum_ib_iu\Big(\ln(\beta_i r_i),x_i\Big) + \lambda(\sum_i\beta_i-1)\,.
\end{equation}
By setting $\partial {\cal L}/\partial\beta_k=0$ we compute
\begin{equation}\label{Eq:AlmostThere}
b_k\,u'\Big(\ln(\beta_k r_k),x_k\Big)=-\lambda\beta_k\,,
\end{equation}
where prime denotes derivative with respect to the first argument. This implies
\begin{equation}
    \sum_i b_i\,u'\Big(\ln(\beta_i r_i),x_i\Big)=-\lambda\sum_i\beta_i=-\lambda\,.
\end{equation}
Substituting this back into Eq.~(\ref{Eq:AlmostThere}) we obtain
\begin{equation}\label{Eq:ExpectedUtilitySolution}
    \beta_k=\frac{u'\Big(\ln(\beta_k r_k),x_k\Big)\,b_k}{\sum_i u'\Big(\ln(\beta_i r_i),x_i\Big)\,b_i}\,.
\end{equation}
The positivity of $\beta_k$ is guaranteed by the monotonicity of the utility function with respect to return $u'>0$ (the limiting case of zero $\beta_k$ is trivial and can be removed from the optimization problem~(\ref{Eq:Lagrangian})). Solving this equation for $\beta_k$ gives us the optimal investment strategy. This includes the growth-optimizing investor as a special case. Indeed, in this case $u'=1$, Eq.~(\ref{Eq:ExpectedUtilitySolution}) reduces to the Kelly equation, $\beta_k=b_k$, which in view of~(\ref{Eq:Bayes_General}), leads to our basic equation~(\ref{Eq:Bayes_GrowthOptimal}).

Research in economics and finance is often done in the comfort of fully continuous settings where the tools of calculus are more readily available. Although our approach does not need that, it would be easier for us to see the emergence of classical concepts if we present our main arguments in a similar fashion. So, substituting Eq.~(\ref{Eq:Fk}) into~(\ref{Eq:ExpectedUtilitySolution}) and writing the result in a fully continuous notation we get\footnote{Discrete treatment is reviewed in Appendix A.}
\begin{equation}\label{Eq:Bayes_UtilityAdjustment}
\beta(x)=\frac{u'\Big(\ln F(x), x\Big)}{\int u'\Big(\ln F(\tilde{x}),\tilde{x}\Big)\,b(\tilde{x})\,d\tilde{x}}\,b(x)\,.
\end{equation}

\begin{center}
\includegraphics[width=0.8\textwidth]{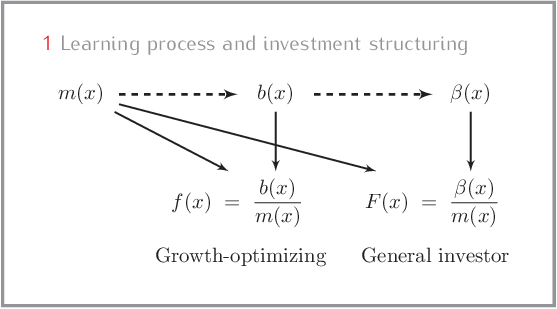}
\end{center}

We are now in a position to understand the logic of how our investor learns and how this learning process translates into investment decisions.

The investor needs to learn two kinds of information. On the one hand, they need to learn about the market. On the other hand, they need to learn about themselves (their preferences and goals). It is a fundamental property of Bayesian learning that, given a set of data, it does not matter in which order the various parts of this data are included in the calculation~\cite{Jaynes_2003}. Indeed, conditional probabilities $P(x|A,B)$ and $P(x|B,A)$ are numerically equal. So we have a lot of freedom in how to organize our calculations.

Great, but how do we start? This question is infamous in statistics as the problem of choosing the prior. In general, choosing the prior can be very hard and the end solution is often ad-hoc. We are lucky, however, to have a natural starting point. This is a growth-optimizing investor who chooses $m(x)$ as the prior information on the market. Why growth-optimizing? Because of the investor equivalence principle. Why $m(x)$? Because, so far, this is all that we have given them -- the prices $\{p_i\}$ of the securities $\{s_i\}$. We emphasize that this is just a starting point of a calculation.

The investor is now in a position to include the rest of their data. They discover related markets, analyze historical records. At this point they may notice differences in drifts implied by their research vs $m(x)$. In some such cases, they may want to speak in terms of \lq\lq real" and \lq\lq risk-neutral" worlds to articulate their findings. The reader has almost certainly witnessed such conversations in practice. Eventually the investor stops researching the markets and writes down their believed distribution~$b(x)$.

A true growth-optimizing investor would normally stop here and proceed directly to the construction of the payoff via equation~(\ref{Eq:Bayes_GrowthOptimal}), as shown on Figure 1. In general, however, the investor might have additional information about their personal preferences. The investor needs to incorporate this information into the calculation given, of course, what they already included about the market. The investor equivalence principle tells us how we can describe this additional learning step mathematically.

Indeed, according to the investor equivalence principle, we can continue to view our investor as growth-optimizing as long as we can imagine persuading them to believe $\beta(x)$. It is important to emphasize that we do not simply imagine someone who already believes $\beta(x)$. We demand logical continuity: we started with a growth-optimizing investor, they have already learned about the market, and now we want them to learn additional information which would make them behave as if they had a different system of preferences. In other words, we want to see if our investor can take their belief $b(x)$ and update it once again using Bayes' theorem arriving at $\beta(x)$. Equation~(\ref{Eq:Bayes_UtilityAdjustment}) describes exactly this additional learning step and Figure 1 summarizes the whole process including the derivation of the final optimal payoff $F$.

The above discussion leads us to two important observations. First, although not entirely obvious from its structure, Eq.~(\ref{Eq:Bayes_UtilityAdjustment}) is about incorporating information on risk aversion into product design. In subsequent sections we further distill this idea and demonstrate its power on practical applications. Second, we see that incorporating {\sl all} information that is relevant to investments ({\sl objective or personal}) is clearly an important ability for an investor. Those investors which can do that without contradicting the logic of Bayes' theorem are especially important, so we give them a name. Let us call them {\sl logical investors}. In the Appendix B we provide additional context both to justify this name and to give a better feel for the role such investors play in our theory.

\section{Payoff elasticity equation}

Dividing both sides of Eq.~(\ref{Eq:Bayes_UtilityAdjustment}) by $m(x)$ and using equations~(\ref{Eq:Bayes_GrowthOptimal}) and~(\ref{Eq:Bayes_General}) we derive
\begin{equation}\label{Eq:EquivalencePrinciple}
F(x)=\frac{u'\Big(\ln F(x),x\Big)}{\int u'\Big(\ln F(\tilde{x}),\tilde{x}\Big)\,b(\tilde{x})\,d\tilde{x}}\,f(x)\,.
\end{equation}
This is an integral equation for $F(x)$. We want to convert it into a more practical differential form.

The concept of elasticity gives us a particular way of differentiating that proved to be especially useful in economics and finance. Elasticity of a function, $\phi(x)$, with respect to its argument, $x$, is defined as the derivative
\begin{equation}
    \frac{d\ln\phi(x)}{d\ln x}\,.
\end{equation}
This measures the percentage change in the function's value with respect to percentage change in its argument.

The classical notion of utility does not depend on $x$ explicitly~\cite{NeumannMorgenstern_1944}. In our notation this is the case when the second argument of $u$ happens to be redundant: $u(\ln F,x)=u(\ln F)$. We investigate this case first and then discuss what happens in general. Taking the logarithm on both sides of Eq.~(\ref{Eq:EquivalencePrinciple}), forgetting temporarily about the explicit dependence of $u$ on $x$ and differentiating we obtain
\begin{equation}
    d\ln F=\frac{1}{u'(\ln F)}\,u''(\ln F)\,d\ln F+d\ln f\,.
\end{equation}
Note that $x$ is the only variable quantity. The denominator on the rhs of Eq.~(\ref{Eq:EquivalencePrinciple}) does not depend on $x$ and drops out from the above differential equation.

Rearranging the terms,
\begin{equation}\label{Eq:PayoffElasticity}
    \frac{d\ln F}{d\ln f}=\frac{u'(\ln F)}{u'(\ln F)-u''(\ln F)}\,.
\end{equation}
Thinking about practical applications, removing the explicit use of the highly theoretical concept of utility from the above equation is our next challenge.

It is well known that affine transformations of the utility function do not have any effect on investors' preferences. Utility functions effectively ignore the two most fundamental mathematical operations -- addition and multiplication by a number. This fact limits practical use of utility functions, especially when we want to talk about risk aversion. To remedy the situation, a variety of measures for risk aversion were introduced. The most popular of them are probably the Arrow-Pratt measures of absolute and relative risk aversion:
\begin{equation}\label{Eq:AP_RiskAversion}
A(F)=-\frac{U''(F)}{U'(F)}\,,\ \ \  R(F)=-F\frac{U''(F)}{U'(F)}\,,
\end{equation}
where $U$ is the standard definition of utility which is connected to our slightly more general notion, $u$, via the equation $U(F(x))=u(\ln F(x))$. By direct calculation we have
\begin{equation}
    U'(F)=\frac{u'(\ln F)}{F}\,,\ \ \ U''(F)=\frac{u''(\ln F)-u'(\ln F)}{F^2}\,.
\end{equation}
Using the definition of relative risk aversion~(\ref{Eq:AP_RiskAversion}), we can now rewrite Eq.~(\ref{Eq:PayoffElasticity}) as
\begin{equation}\label{Eq:ThePayoffElasticityEquation}
    \frac{d\ln F}{d\ln f}=\frac{1}{R}\,.
\end{equation}
This simple equation is the central technical result of this paper. It gives us a fundamental link between payoff elasticity and risk aversion. The more risk aversion we have the less elastic is the payoff.

On the practical side, this equation allows us to compute the optimal payoff $F$ from the growth-optimal $f$ and the risk aversion profile $R$ of the client. Conversely, we are now also able to compute risk aversion profiles directly from clients' positions.

For completeness, we mention some alternative forms of Eq.~(\ref{Eq:ThePayoffElasticityEquation}) which can be used depending on the application. Considering the payoff elasticity equation~(\ref{Eq:ThePayoffElasticityEquation}) for two different general investors we derive
 \begin{equation}
\frac{d\ln F^{(1)}}{d\ln F^{(2)}}=\frac{R^{(2)}}{R^{(1)}}.
\end{equation}
This equation shows that other payoff profiles (not necessarily growth-optimal) can serve us as building blocks. The concept of elasticity can be replaced by the ordinary differentiation if we decide to work in terms of a less fundamental measure of absolute risk aversion:
 \begin{equation}\label{Eq:PayoffDerivativeEquations}
\frac{dF}{df}=\frac{1}{fA}\,,\ \ \ \frac{dF^{(1)}}{dF^{(2)}}=\frac{A^{(2)}}{A^{(1)}}\,.
\end{equation}
How would the above derivation change if we allowed for explicit dependence of $u$ on $x$? The payoff elasticity equation would stay the same but the expressions for both $R$ and $A$ would become more complicated. In particular, they would acquire explicit dependence on the state $x$ (see Appendix A), but would of course reduce to the original Arrow-Pratt definitions in the special case of state-independent preferences. On the practical level, the important thing to mention is that essentially all of these equations have been solved and present no further technical challenge. Indeed, the Picard-Lindel\"of theorem provides the necessary theory (for the one-dimensional case) and even offers an explicit construction of the general solution (Picard iteration method).

\section{Illustrations}
In this section we illustrate practical usage of the payoff elasticity equation. We warm up on a simple analytically solvable example and show how to examine structuring ideas that do not necessarily come from our theory. We then turn to the main power of the payoff elasticity equation -- the ability to understand and to adjust clients' risk aversion. We show that naive attempts to express risk aversion can be dangerous and that the payoff elasticity equation
provides us with a more sound technology.

\subsection{One-parameter investor families}
By a one-parameter family we mean a set of investors whose degree of risk aversion is controlled by a single number, e.g. constant absolute or constant relative risk aversion. The usefulness of one-parameter families comes from the fact that the position of an investor within the family can be determined by asking the investor a single question regarding, for instance, their maximum acceptable loss. To illustrate this point, let us look at Eq.~(\ref{Eq:PayoffDerivativeEquations}) and consider the case when it takes a particularly simple form: we set
\begin{equation}
    A=\frac{a}{f}\,,
\end{equation}
where $a$ is a constant which controls the strength of risk aversion. We immediately derive
\begin{equation}\label{Eq:SimplestInvestor}
    F=\frac{1}{a}(f-1)+1\,.
\end{equation}
We see that $a$ effectively scales the payoff around the bond line $F=1$. This is exactly how Figure 2 in~\cite{Soklakov_2011} was constructed. The value of $a$ can be easily found by matching investors maximum acceptable loss.

\subsection{Product validation}
Can our theory accommodate every imaginable investor? Of course not. One can imagine investors that demand greater flexibility than our equations would allow. As explained in Appendix B, we should be very cautious in offering assistance to such investors as not every desire is necessarily wise. At the moment it may be more prudent to focus on the many unexplored possibilities that are already offered by our theory.

Having said that, it is very important to acknowledge that investment ideas are born in all sorts of ways and good ideas may not necessarily come through an explicit use of our theory. To recognize such cases we want the ability to verify that a given investment strategy is both rational and logical.

In order to see how this can be done, let us examine the family of investors considered in~\cite{Shimko_1994}. The investors are looking for a payoff structure, $h(x)$, which solves a certain optimization problem. Using our notation, the optimization problem reads:
\begin{eqnarray}
    \max_{h}\Big[\underbrace{\int h(x) b(x)\,dx}_{\rm mean}-\frac{R_a}{2}\underbrace{\int h^2(x) m(x)\, dx}_{\rm variance}\Big] \label{max:Shimko}\\
    {\rm subject\ to\ \ }\int h(x) m(x) \, dx=0\,, \label{constraint:Shimko}
\end{eqnarray}
where $R_a$ is a parameter that controls the degree of risk aversion. In Ref.~\cite{Shimko_1994} this problem was introduced by analogy with the mean-variance optimization approach of Markowitz~\cite{Markowitz_1952}. It is important to note, however, that the mean and the variance, given by the first and the second term of~(\ref{max:Shimko}) respectively, are computed using two very different distributions. In particular, the mean is computed using the investor-believed $b(x)$, while the definition of variance is using the market-implied $m(x)$. As a result, the optimization setup~(\ref{max:Shimko}-\ref{constraint:Shimko}) does not fit well into utility maximization paradigm, so, on its own, the setup~(\ref{max:Shimko}-\ref{constraint:Shimko}) is not enough to recommend the investment as rational. The idea of penalizing market-based variance does, however, make significant intuitive sense, so let us not rush into dismissing it on formal grounds. Let us independently investigate in what sense the above idea may point towards a sensible product.

In what follows we put aside any attempts to justify the optimization~(\ref{max:Shimko}-\ref{constraint:Shimko}) and proceed by directly examining its solution, $h(x)$. By doing so we put ourselves in a rather typical situation when an investment product is proposed which looks good yet with some questions regarding its real quality. Before we can recommend the product, we want to verify the assumptions under which $h(x)$ may be viewed as a rational strategy pursued by a logical investor. The expression for $h(x)$ is given by the first equation in~\cite{Shimko_1994}, namely
\begin{equation}
    h(x)=\frac{b(x)-m(x)}{m(x)}\frac{1}{R_a}\,.
\end{equation}
By rearranging the terms we can rewrite this equation as
\begin{equation}
    b(x)=\Big(R_a h(x)+1\Big) m(x)\,.
\end{equation}
This allows us to understand the investment through the eyes of a growth-optimizing investor. Indeed, using equation~(\ref{Eq:Bayes_GrowthOptimal}) we derive
\begin{equation}
    R_a h(x)+1=f(x)\,,
\end{equation}
where $f(x)$ is the growth-optimal payoff structure. Finally, we see that
\begin{equation}\label{Eq:h_understood}
    h(x)=\frac{1}{R_a}\Big(f(x)-1\Big)\,.
\end{equation}
We immediately recognize this equation as a particular case of the one-parameter investor family which we considered above. Indeed, the only difference between this equation and equation~(\ref{Eq:SimplestInvestor}) is the absence of an additive constant -- the upfront investment cost of 1 -- which drops out from (\ref{Eq:h_understood}) on the account of the constraint~(\ref{constraint:Shimko}). This constraint assumes the existence of a perfectly liquid two-way market in options on $x$ which effectively allows the investor to borrow from the market and set up the investment at zero upfront cost. Such assumptions are very important to note and check. Other than that we managed to verify that the solution of the optimization~(\ref{max:Shimko}-\ref{constraint:Shimko}) can be viewed as a rational investment pursued by a logical investor with an understandable risk aversion strategy.

\subsection{Implying and adjusting risk aversion}

Empirical computations of risk aversion have always been a very challenging task. We now have additional tools. We can imply risk aversion directly from clients' positions or ideas, adjust it if necessary and use it to structure the optimal product.

To emphasize the immediate relevance of our techniques in the context of current practices, let us review one of the most popular ideas in finance -- expressing higher risk aversion by implementing a more conservative view. This very simple idea is not confined to finance. It almost certainly came into our field from other more basic human behaviors. It obviously has powerful evolutionary reasons yet it is not without faults. Failing strategies, later described as \lq\lq half-measures" or \lq\lq indecisions", often begin as conservative interpretations of data. So let us examine how robust this idea really is in the mathematical context of financial structuring.

Imagine a growth-optimizing investor which believes that the skew should be half of the value observed on the market. This view and the corresponding optimal payout are depicted in red on Figures 2.A and 2.B. The investor decides to be more cautious and to avoid expressing their view in the wings. To this end they manually adjusts their original view by gently forcing it to revert back to the market in the wings (blue line on Figure 2.A). The corresponding payoff structure is given by the blue line on Figure 2.B.

We can now use the payoff elasticity equation and examine the risk aversion profile which corresponds to investor's attempt to soften their view. Before we do that, let us think what we expect to see. Risk aversion profile for a growth-optimizing investor is flat $R=1$. So, surely, what we should see is a profile that lies above the line $R=1$ increasing further in the wings.

Looking at Figure 2.C we see that this intuition could not be further from the truth. Not only the profile does not form a convex line above growth-optimizing, it widely oscillates crossing $R=1$ and even goes negative. Was the investor too crude in modifying their view? Were they too harsh or too lenient? Either way, how could they become less risk averse and even crossed the line into risk-loving territory of gambling? We encourage the reader to try their own ways of modifying views. The more you try the more evidence you obtain supporting the inevitable conclusion -- it is very difficult, practically impossible, to achieve consistent sensible risk aversion by ad-hoc modification of views.
\begin{center}
\includegraphics[width=\textwidth]{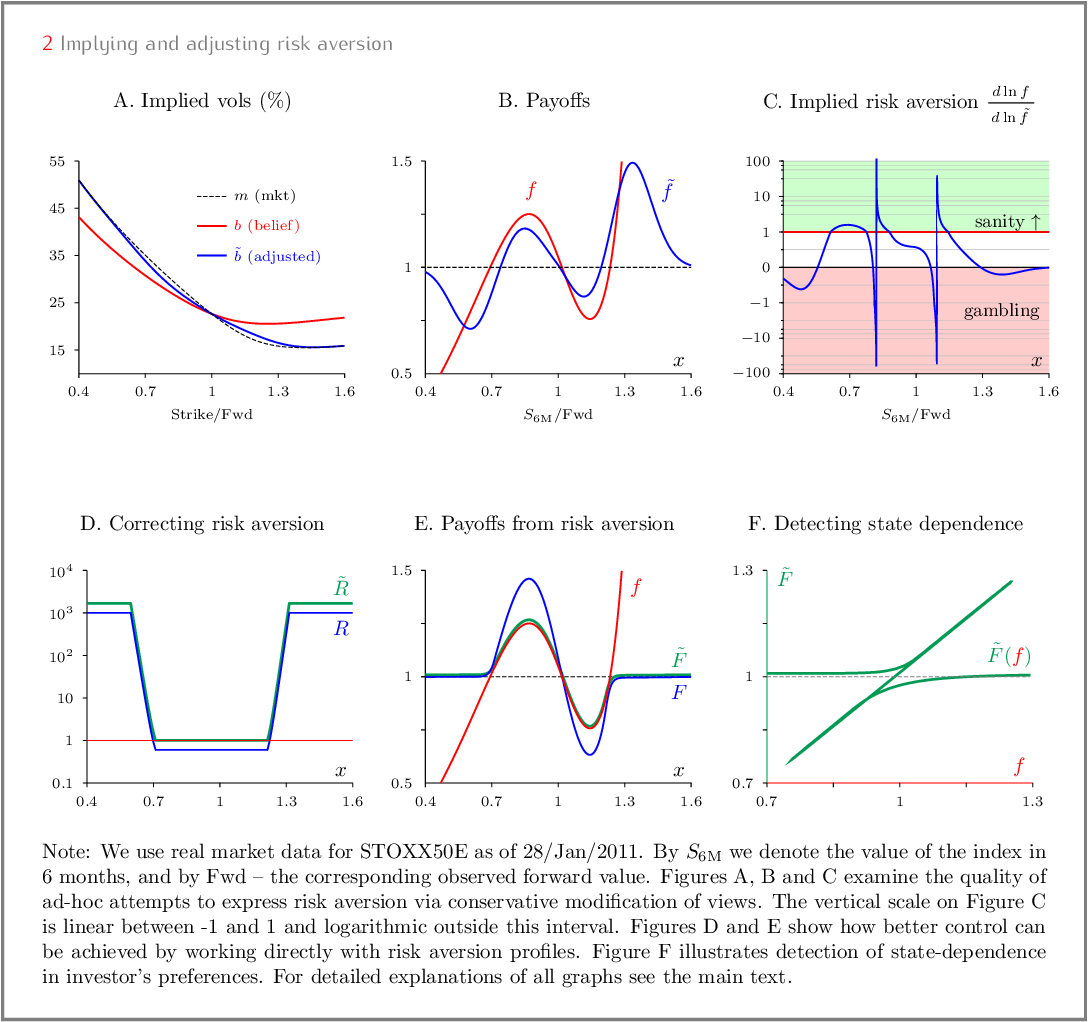}
\end{center}

Using the payoff elasticity equation gives us a much easier way of including risk aversion. To demonstrate let us create a simple textbook example. We take the same market and the same growth-optimizing investor as we considered above. This time, however, we accommodate the investor's desire to suppress their view in the wings by stating a very high level of risk aversion -- see Figure 2.D. Integrating the payoff elasticity equation immediately gives us the payoff structure -- Figure 2.E.

The two Figures 2.D and 2.E also illustrate the corrective effect of adjusting risk aversion profiles. By comparison to growth-optimizing, the investor with the risk aversion $R$ is more risk-averse in the wings ($R>1$) and less risk-averse near ATM ($R<1$). Looking at the payoff (blue line $F$ on Figure 2.E vs the red line $f$) we see that the investor effectively reallocates their exposure from the wings to the ATM region. This reallocation effect can be perfectly removed (if desired) by replacing $R$ with $\tilde{R}$, i.e. by making sure that we are at least as risk averse as the growth-optimizing investor. Having obtained the payoffs we can easily illustrate them in terms of market views. This would give us pictures which resemble Figure 2.A in the general shape with the difference that this time we don't just hope (and almost surely fail) to imply some reasonable risk aversion -- we know exactly what we are building (this is how Investor 2 on Figure 2.A in Ref~\cite{Soklakov_2013a} was constructed).

Direct modifications of payoffs is yet another example of popular structuring ad-hocery for imitating risk aversion. The most benign examples of that are introducing ad-hoc floors and caps on top of otherwise problematic payoffs. To see that direct payoff modifications can be dangerous, all we have to do is to examine modifications of $f$ presented on Figures 2.B and 2.E. At first glance all of them look reasonable.

As the above examples illustrate, modifying payoff structures by changing risk aversion profiles is easy and intuitive. With almost no practice you will see how to produce structures that appear to have caps, floors and other recognizable features that can be very useful in describing the resulting product. As before, we don't just hope to produce sensible risk aversion as a byproduct of ad-hoc modifications -- we know exactly what risk aversion profile we are using.

\subsection{Rules of thumb}

Handling risk aversion is a complicated and delicate enough task to require a specialist tool. Payoff elasticity equation gives us such a tool. It would be even better, however, if we could say something practical about products even when knowing next to nothing about risk preferences. Amazingly enough, this is possible.

Note, for instance, that most investors would want to express positive risk aversion.\footnote{This observation comes from understanding the difference between risk-averse investors and risk-loving gamblers.} We immediately see that positive risk aversion implies that $F$ and $f$ must agree on the location and the type of their extremum points. Indeed, as we vary $x$ both $F(x)$ and $f(x)$ should be going up or down together for the elasticity $d\ln F/d\ln f$ to remain positive. Violation of this simple rule is what produced the wildly oscillating often negative profile of Figure~2.C.

We can talk about state-agnostic and state-dependent investors. For the former kind the utility of their investment depends only on its performance -- the classical Arrow-Pratt investor. The latter kind also cares about the underlying market at maturity. Clients would normally know which kind they are and we can easily test if they get the right kind of product. Indeed, for a state-agnostic investor (imagine a US investor who is exposed to UK market only via their investment), $R$ depends on $x$ only via $F$. In this case, by the payoff elasticity equation, $F$ becomes a function of $f$. This we can test by simply plotting $F$ against $f$ (while varying $x$ as a parameter) and then examining the picture. Figure~2.F, for instance, shows that $\tilde F$ cannot be computed from $f$ alone -- a typical signature of a state-dependent investment.

We learned that the locations of the maxima and minima of $F$ are essentially determined by that of $f$. Given that $F$ is just a function, the locations and the types of its extremal points already give us a lot of information. A lot more can be said about $F$ which belongs to a state-agnostic investor. The fact that $F$ depends on $x$ only via $f$ imposes severe constraints. We can see, for example, that $F$ and $f$ intersect on a bond line, i.e. if $F(x_1)=f(x_1)$ and $F(x_2)=f(x_2)$ then $F(x_1)=F(x_2)=c$, where $c$ is a constant (not necessarily 1).

\includegraphics[width=\textwidth]{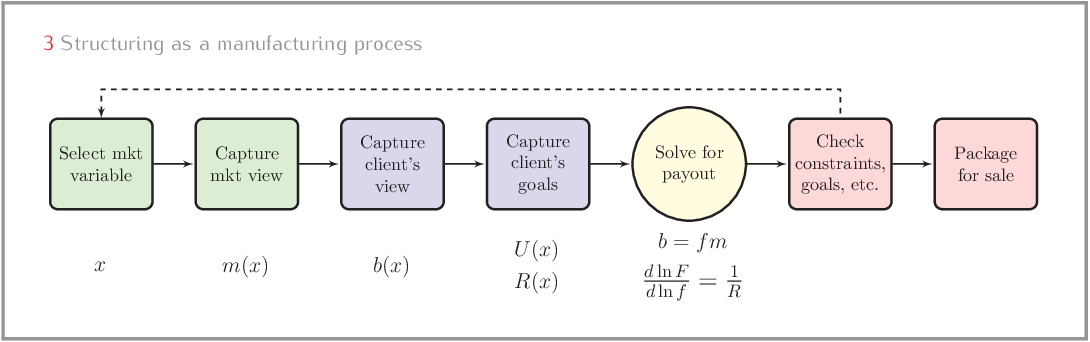}\\

Because structuring is often constrained by practical considerations (such as the number of traded strikes which we could use for replicating the final product) such rules of thumb provide us with a lot of information. Couple that with additional data like maximum tolerable loss and you might be able to sketch the final optimal structure without detailed knowledge of your client's risk aversion.

\section{Summary}

At the heart of structuring we always have optimization which reflects the goals of the client. This is preceded by preparatory steps defining the problem and followed by packaging of the solution into a tradeable product. Together these steps form the backbone of a manufacturing process which we summarized on Figure 3.

The preparatory steps include deciding on the market variable, capturing the relevant market and client views and identifying the goals of the client. This leads us to the key solution stage where our equations come in: the growth-optimal $b=fm$ and the payoff elasticity equation.

After the solution stage we check the quality of the derived product. At this point we might be interested in assessing the expected performance of the product. The relevant quantitative techniques are explained and illustrated in Refs.~\cite{Soklakov_2014MR} and~\cite{Soklakov_2014EqPuzzle}. We might even decide to go back and redefine the underlying variable. In~\cite{Soklakov_2013a} we considered a detailed example when this happens. Having satisfied ourselves with the solution we proceed to the final stage of packaging it into a tradeable product.

\section{Appendix A: Discrete elasticity equation}
In the derivation of the payoff elasticity equation we made two assumptions: first, we considered continuous $x$ and, second, we considered state-independent utility functions, i.e. utilities which depend on $x$ only via the value of the payoff $F(x)$. We did it simply to highlight the connection with the classic Arrow-Pratt measures of risk aversion which were built on these assumptions. In the actual fact, none of these assumptions is important to us except, of course, that the continuous case benefits from the rich toolbox of calculus.

\newpage
A few lines of algebra give us a discrete fully general payoff elasticity equation. Indeed, dividing both sides of Eq.~(\ref{Eq:ExpectedUtilitySolution}) by $m_k$ and using Eqs.~(\ref{Eq:Fk}), (\ref{Eq:Bayes_General}) and~(\ref{Eq:Bayes_GrowthOptimal}) we obtain
\begin{equation}
    F_k=\frac{u'\Big(\ln F_k,x_k\Big)\,f_k}{\sum_i u'\Big(\ln F_i,x_i\Big)\,b_i}\,.
\end{equation}
We then compute
\begin{equation}
\frac{F_{k+1}}{F_k}=\frac{u'\Big(\ln F_{k+1},x_{k+1}\Big)\,f_{k+1}}{u'\Big(\ln F_k,x_k\Big)\,f_k}\,.
\end{equation}
Taking logarithm on both sides and rearranging terms,
\begin{equation}
\ln\frac{F_{k+1}}{F_k}\cdot\left(1-\frac{\ln\frac{u'(\ln F_{k+1},x_{k+1})}{u'(\ln F_k,x_k)}}{\ln\frac{F_{k+1}}{F_k}}\right)=\ln\frac{f_{k+1}}{f_k}\,.
\end{equation}
Finally we get a discrete payoff elasticity equation
\begin{equation}
\frac{\ln\frac{F_{k+1}}{F_k}}{\ln\frac{f_{k+1}}{f_k}}=\frac{1}{R_{k,k+1}}\,,
\end{equation}
where the expression for $R_{k,k+1}$ in terms of utility can be easily obtained from the previous equation. We see that, in contrast to the original Arrow-Pratt notion, $R_{k,k+1}$ depends not only on $F$ but also on $x$. Detailed investigations of state-dependent utilities and the emergence of $R_{k,k+1}$ as a generalization of the Arrow-Pratt risk aversion are very interesting topics. These topics, however, would take us far beyond the scope of this~paper.

\section{Appendix B: Rationality and Logic}

Together with rationality, as captured by the expected utility approach, there is another technical aspect of decision making which is extremely important but which is often taken for granted. This aspect is the mathematical foundation of logic itself.

The simplest and perhaps the most widely known example of logical framework is given by the Boolean algebra which uses True=1 and False=0 to describe logical statements. Boolean algebra has proven itself as an extremely powerful tool underpinning, among other things, all computer-based calculations. Amazingly enough, despite all of its successes, the Boolean approach leaves substantial room for improvement.

Indeed, what should we do with statements for which we cannot say if they are true or false? Imagine, for example, a statement for which we only have a measure of how often or how likely it is to be true. It turns out that the Boolean approach can be very easily generalized to handle such cases~\cite{Jaynes_2003}. All we have to do is to replace True and False by numbers between zero and one and, of course, we have to generalize the rules for manipulating these numbers. The resulting logical system turns out to coincide with the standard probability theory. In particular, the product rule which we use to compute joint probability, i.e.
\begin{equation}\label{Eq:BayesFromLogic}
p(A,B)=p(A)p(B|A)=p(B)p(A|B)\,,
\end{equation}
arises as a generalization of the Boolean product which we use to determine the truth value of the joint statement $(A {\rm\ and\ }B)$. The only conceptual difference is that statements can now have varying degree of information about one another which is handled by the concept of conditioning (as in conditional probabilities). The last equality in~(\ref{Eq:BayesFromLogic}), i.e. Bayes' theorem, emerges as a logical consistency requirement which comes from swapping the order in which we include $A$ and $B$ into the calculation.

This intimate connection of Bayes' theorem with logic underlies the emphasis we make on logical investors. Indeed, violations of Bayes' theorem do not just go against basic probability theory. Such violations may also be a symptom of much deeper logical inconsistencies. It is our duty to help our investors to remain logical. Of course, the utility theory itself is already based on logic. We can therefore expect that, at least partially, logical behavior is already included in rationality. This, I think, is what allowed us to use such simple derivations.

For a comprehensive literature review of the Bayesian methods in portfolio selection and market risk management we point to~\cite{RachevEtAl_2008} and references therein. For example applications outside market risk (e.g. credit risk and stress testing) see Refs.~\cite{JacobsKiefer_2010} and~\cite{JacobsEtAl_2015}.


\end{document}